\documentclass[12pt]{article}
\usepackage{graphicx}                                                                                 
\begin{document} \begin{center}
                                                                                
{\huge Crack Patterns in Laponite Films Dried in Electrostatic Field }\\
\vskip .5cm
 Sujata Tarafdar, Dibyendu Mal, Suparna Sinha, and T R Middya \\ \vskip .5cm
Condensed Matter Physics Research Centre, Physics Department\\ Jadavpur University, Kolkata 700032\\India
\end{center}
\vskip 1cm
\noindent {\bf Abstract}\\
Crack patterns in a layer of laponite gel allowed to dry in a static electric field are studied.
Crack patterns in natural and synthetic clays have been studied extensively with interesting results. 
This is the first report of such an observation in a radial electric field.
The nano-sized disc-like laponite particles carry a quadrupole moment due to their charge distribution. The
interaction of the quadrupole moment with the field gradient in a non-uniform field of radial symmetry is
probably responsible for the characteristic pattern observed. The cracks start radially from the positive
electrode. The same geometry with no field does not produce the characteristic pattern, neither does a uniform field
with rectangular geometry.
\vskip 1cm
   
Desiccation cracks in clay materials make an interesting study which may be of practical importance \cite{dm,frac,
crack2, crack3}. Clay particles are flat and plate-like usually of micron size with a 
surface charge \cite{clay}, so they are likely to be 
influenced by an external electric field. 
It is more convenient in laboratory experiments to study clean chemistry customized synthetic clays like laponite RD.
We let films of laponite gel dry in a static electric field, the pattern
of desiccation cracks formed is found to be strongly affected by the geometry and the the direction of the electric 
field gradient. A radial electric field with cylindrical symmetry produces a radial pattern of cracks, distinctly different
from a drying gel in the same geometrical set-up but with no field. A uniform electric field in ractangular geometry
is also found to have no effect. Moreover, reversing the direction of the field
reverses the pattern, the cracks always starting from the positive terminal. It is possible to interpret the origin 
of the pattern
from the known charge distribution on the laponite platelets \cite{foss1}. 

The experimental procedure is as follows. 2.5 gm. laponite RD (Rockwood Additives) is mixed with 50 ml. 
distilled water. The mixture is stirred for 15 minutes in a magnetic stirrer and deposited in
circular petri dishes of 12 cm diameter and allowed to dry. The thick suspension just before formation of the gel
is mildly alkaline with a pH of 9.5. Two electrodes constructed from aluminium foil are fitted
to the petri dishes. One is in the form of a thin rod placed at the centre of the dish and the counter-electrode consists
of an aluminium strip placed at the edge of the petri-dish in the form of a short cylinder. A static field is applied 
from a constant voltage power supply, between the two electrodes. For comparison we dry a set of samples in 
an identical arrangement but without the applied voltage.

\begin{figure}[h]
\begin{center}
\includegraphics[width=9.0cm]{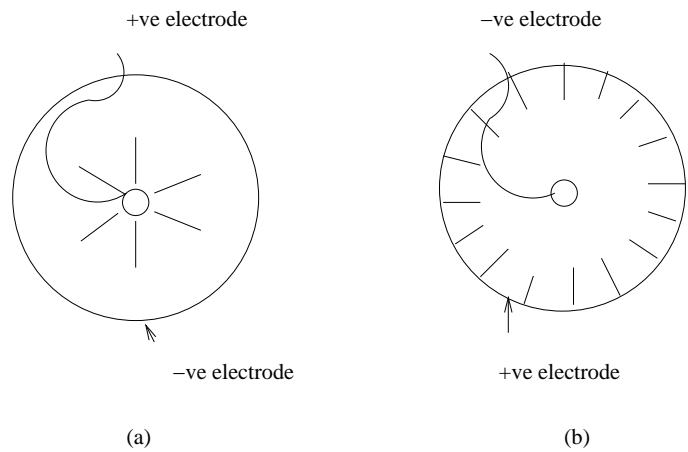}
\end{center}
\caption{A schematic diagram showing the crack development for two experimental set-ups, (a) with the centre
electrode positive and (b) with the central electrode negative.  } \label{diag}
\end{figure}

A schematic picture figure(\ref{diag}) shows the geometry of the set-up. We find that radial crack patterns start
developing earlier in the samples with applied field, than in the sample without field. With the positive terminal in
the centre, the cracks appear at the central electrode and move out radially in a very straight and symmetric array.

\begin{figure}[h]
\begin{center}
\includegraphics[width=7.0cm, angle=270]{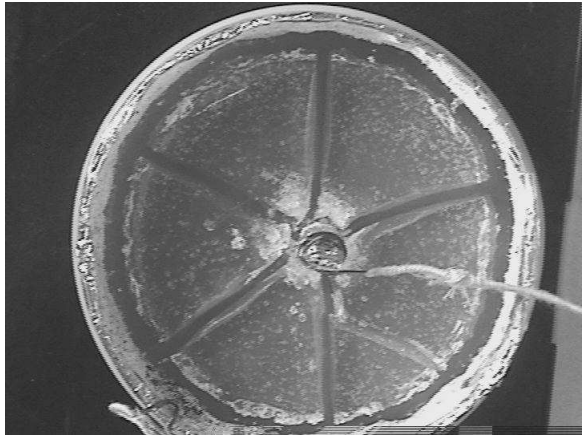}
\end{center}
\caption{Crack patterns developed for a field of 50 volts with the center terminal positive. 
}
\label{pos1}
\end{figure}
                                                                                                                             
This is shown in the photograph in Figure(\ref{pos1}).
In the 
samples without field, the pattern is a network  of cracks forming rectangular peds as 
normally observed \cite{crack2}. Here the gel adheres to the central electrode and the cracks never meet it, unlike the
case with the static field. If the central electrode is attached to the negative terminal of the power supply, 
the cracks appear at the outer
boundary of the petri-dish and radiate inwards. However, the pattern is less regular in this case, and disordered
cracks start forming earlier. The static field
has been varied from 0 - 200 V, we find that the pattern starts to lose its distinctive strictly radial nature
for a field strength below 50 V. 

While the 50 V pattern  shows cracks emanating almost perfectly in the 
radial direction from the
centre, in the 30 V pattern the cracks deviate considerably from the radial direction. 

With a  set-up in where the centre terminaL has opposite polarity, The cracks again form radially from 
the positive end, in this case the periphery, and proceed inward.
However, here other disordered cracks start forming before the peripheral cracks reach the centre. This is 
probably because the positive field at the boundary is weaker and the field-induced cracks take more time to form,
in the mean time the sample dries further and the field effect gets screened by the large air gaps formed near the
electrodes.

To see whether the field {\it gradient} is responsible for the pattern, several samples were deposited in rectangular
shallow troughs with positive and negative electrodes on opposite sides. Here the pattern shows no effect of the
electric field.

An explanation for the observed patterns may be offered as follows. Laponite is a synthetic clay with the chemical
formula ${Na_{0.7}}^+[(Si_8Mg_{5.5}Li_{0.4})O_{20}(OH)_4]^{0.7-}$, it consists of monodisperse flat plate-like 
particles of about 25 nm diameter and 1 nm
thickness. When dispersed in water, the plates acquire a net negative charge of several hundred electron charge (-e)
The flat surface of the plates are negatively charged, while the rims have a positive charge. The total positive
charge on the rim is approximately some tens of e, much less in magnitude than the negative charge. Due to the 
opposite charges on the flat surface and rims
laponite is supposed to form aggregates with a house-of-cards structure \cite{lap}. The aggregate is built up of mainly 
T-shaped units \cite{dijk}.

\begin{figure}[h]
\begin{center}
\includegraphics[width=8.0cm, angle=270]{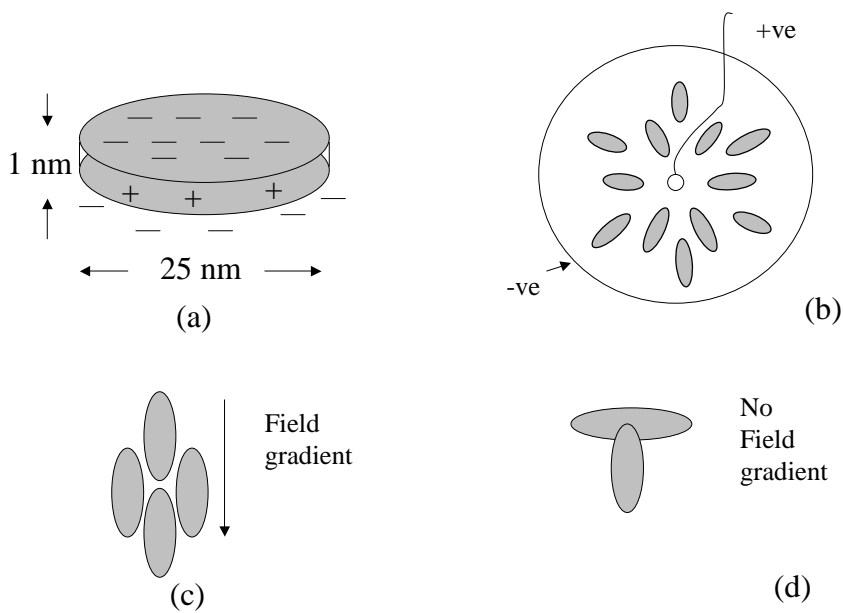}
\end{center}
\caption{(a) A laponite disc showing surface charge distribution, (b) suggested alignment of laponite discs in the radial 
field. (c) Position of adjacent discs in an electric field gradient due to their quadrupole moment,
(d) The T-configuration for discs with uniform or zero field. The ellipses show end-on positions of the laponite discs.}
\label{discs}
\end{figure}

The laponite discs, due to their chemical composition have the structure of a symmetrically sandwiched 
positively charged layer between
two negative layers, has a C${_{2v}}$ point group symmetry. The charge distribution has a net negative monopole
charge, zero dipole moment and a non-zero quadrupole moment. The monopole charge will draw the disc as a whole towards
the positive side in an electric field. If the field is non-uniform, as in the radial geometry case, the field
gradient,  interacts with the quadrupole moment, causing the disc to align in the radial direction.                                                                               
Figure(\ref{discs}) shows a schematic diagram of the only possible arrangement of the laponite discs which conforms
to the cylindrical symmetry of the set-up. The adjacent discs have the positively charged rim of one against the 
negatively charged flat surface of the other. The discs are more crowded at the positive side the electric field
and hence on desiccation the cracks start forming here. As desiccation starts, the competition between the shrinking 
of the layer and the adhesion of the lower surface to the substrate, causes cracks to appear \cite{crack}. 
In the present arrangement, faces of the discs will move apart due to repulsion between
their similar charges and formation of  radial cracks will be preferred to other patterns. 
The radial orienting of the discs in the field is
due to the interaction between the quadrupole moment of the discs and the gradient of the non-uniform electric field
\cite{qpole}.
In the absence of the field the preferred configurations of the laponite discs are the end-to-face T-shaped units
\cite{dijk}. 

Crack formation in clay is important in practical situations. Laponite, is used commercially in many applications
such as paints, cosmetics, coatings etc. so crack formation in laponite is also an important study.

Laponite \cite{lap} has been extensively studied since its composition is exactly known and it consists of
 monodisperse nanosized platelets. It is a synthetic clay, having the properties of interest of
clays, but free from the heterogeneity of natural clays. Gel and glass formation in laponite
has been investigated by optical methods \cite{gel1,gel2}. It has been found to be affected by
an electric field \cite{teys,fossum}, but this is the first observation of cracking patterns 
under the influence of an electric field. Crack patterns in laponite suspended in methanol
have been studied. Methanol being a non-polar solvent, does not dissolve the laponite and
a fine array of hierarchical cracks are formed \cite{dm,frac}. The cracks in the aqueous
laponite gel are much wider, and have smoother and straighter edges. the laponite gel swells
in water, so much larger samples would have to be prepared to observe the scaling of crack
area, which was observed in the drying methanol-laponite mixture \cite{frac}.

SEM and XRD studies of the samples is in progress to determine the effect of the electric field on the morphology of
the samples at microscopic scales. We also plan to calculate the energy due to the quadrupole moment and field gradient
interaction to get a more quantitative picture.  

{\bf Acknowledgement:} Authors are very grateful to Rockwood Additives for gifting samples of Laponite RD. 
SS is grateful to DST for financial assistance. Authors thank Tapati Dutta and H. van Damme for encouragement and
discussion and N. K. Chattopadhyay for help and valuable suggestions.

\end{document}